\documentclass[aps,pra,superscriptaddress,amsmath,amssymb,preprintnumbers,floatfix,showpacs,12pt]{revtex4}

\usepackage{amssymb}
\usepackage{epsfig}

\begin{document}
\title{ Quantum properties of the codirectional three-mode Kerr nonlinear
coupler }

\author{ Faisal A. A. El-Orany}
\affiliation{ Department of Mathematics  and Computer Science,
Faculty of Science, Suez Canal University,
 Ismailia, Egypt;
Department of Optics, Palack\'{y} University, 17. listopadu 50,
 772 07 Olomouc, Czech Republic}

\author{ M. Sebawe Abdalla}
\affiliation{Mathematics Department, College of Science, King Saud
University, P.O. Box 2455, Riyadh 11451, Saudi Arabia}

\author{ and J. Pe\v {r}ina}
\affiliation{Department of Optics and Joint Laboratory of Optics,
Palack\'{y} University, 17. listopadu 50,
 772 07 Olomouc, Czech Republic}

\date{\today}

\begin{abstract}
We investigate the quantum properties for  the codirectional
three-mode  Kerr nonlinear coupler. We  investigate single-, two-
and three-mode quadrature squeezing, Wigner function and purity.
We prove that this device can provide richer nonclassical effects
 than those produced by the conventional coupler, i.e. the
two-mode Kerr coupler. We show that it can provide squeezing and
the quadrature squeezing exhibiting leaf-revival-collapse
phenomenon in dependence  on the values of the interaction
parameters. In contrast to the conventional Kerr coupler two
different forms of cat states can be simultaneously generated in
the waveguides. We deduce  conditions required for the complete
disentanglement between the components of the system.

\end{abstract}

 \maketitle


\section{Introduction}

The optical coupler is a device composed of two (or more)
waveguides, which are placed close enough to allow exchanging
energy between waveguides via evanescent waves \cite{jen1}.
Recently, this device has attracted much attention for several
reasons. The progress in the optics communication and quantum
computing networks requires data transmission \cite{EKer1}. This
simple device has potential applications in all-optical switching
\cite{alo,{qu10}}. Furthermore, it provides electromagnetic fields
with an exceptionally wide range of nonclassical effects. Most
importantly this device has been implemented
\cite{exp2,{exp1}} and applied in many
experimental approaches, e.g. in  picosecond switching induced by
saturable absorption \cite{Fin}, optical multi-mode
   interference devices based on self-imaging \cite{sold} and
  photonic bandgap structures in planar nonlinear waveguides \cite{tric}.
Also the generation of correlated
   photons in controlled spatial modes by downconversion in
   nonlinear waveguides has been discussed in \cite{bana}.

Quantum mechanically, there are different types of directional
couplers. For instance, symmetric coupler \cite{sym} (linear (or
nonlinear) processes are involved in  both the
 waveguides), asymmetric coupler \cite{asym} (at least one of the waveguides
possesses different nonlinearity than the others), Raman-Brillouin
 coupler \cite{ram}, bandgap coupler \cite{ban} and Kerr coupler
 \cite{qu14,{qu15},{qu18},{faisal1},{ar1},{ar2},{faisal2},{faisal3}}.
For more details about their properties the reader can consult the
review papers \cite{qu20}.

Kerr nonlinear coupler (KNC) has taken a considerable interest in
the literature since the third-order nonlinearity provides an
effective mechanism for generating nonclassical effects in the
electromagnetic waves caused by the  processes of self-phase and
cross-phase modulations. Also this is related to potential
advantage of the possible observation of the large values of
third-order optical nonlinearities in the organic polymers
\cite{exp2}. In this regard  the generation of continous variable
Einstein-Podolsky-Rosen entanglement via Kerr nonlinearity in an
optical fiber \cite{chh} as well as the generation of spatial
soliton arrays in a planar Kerr waveguide from seeded spontaneous
 parametric down conversion \cite{fanj} are achieved.
  Several papers have been
devoted to the quantum properties of the KNC in the framework of
rotating-wave approximation by neglecting the rotational terms in
such a way that a closed form solution is obtained
\cite{qu14,{qu15},{qu18},{faisal1},{ar1},{ar2}}. In this case KNC
has provided many of interesting effects such as revival-collapse
phenomenon (RCP) in the mean-photon numbers, squeezing of vacuum
fluctuations, sub-Poissonian statistics in single as well as in
the compound modes \cite{qu14}. The phase distribution of KNC has
been investigated showing that the phase-difference evolution is
closely connected with the energy exchange between waveguides and
the RCP in the mean-photon numbers is due to the bifurcation of
the phase-difference probability distribution \cite{qu18}.
Furthermore, the geometry of the waveguides  has been considered
via varying linear coupling coefficients in the codirectional  KNC
\cite{qu15} and contradirectional  KNC \cite{ar1,{ar2}}. In these
cases it has been shown that there is a possibility to control the
switching characteristics and principal squeezing effect by
adjusting the shape of the waveguides. Quite recently, we have
investigated the single-mode quantum properties of the
codirectional Kerr nonlinear coupler when the frequency mismatch
is involved and a condition for obtaining an exact solution for
the equations of motion is fulfilled \cite{faisal2,{faisal3}}. For
this case we have shown that the mean-photon numbers exhibit
oscillatory behaviour rather than revivals and collapses.
Additionally  we have proved that the Schr\"{o}dinger-cat states,
in particular, Yurke-Stoler cat states (YSCS) \cite{yur} can be generated.
Also the higher-order squeezing has been investigated
\cite{faisal3}.

Till now the KNC has been treated as a two-mode device. In the
present paper we give for the first time--as far as we know--the
three-mode version of this device. The
motivation of developing such  device is  that the three-mode KNC
can provide nonclassical effects richer than those obtained from
the conventional coupler (i.e. the two-mode version), as we shall
show throughout the paper. As is well known that  the basic
efforts  in the quantum optics is to enhance the nonclassical
effects. Additionally, we show that  the quadrature squeezing
exhibits leaf-revival-collapse phenomenon in dependence on the
values of the interaction parameters.
In this phenomenon the revival patterns provide leaf shapes and between
two revival patterns short collapse period occurs. It is worth reminding
that for the standard revival-collapse phenomenon the revival patterns
have ellipsoid shape.
We proceed that, in contrast to the
conventional Kerr coupler two different forms of cat states can be
simultaneously generated in the waveguides.
 The  investigation of the three-mode KNC will be given in the following order:
In section $2$ we give the Hamiltonian for the system and the
solution for the equations of motion. Also we derive the
expectation values for  different moments of operators, which will
be used in the paper. In section $3$ we investigate the
quadrature squeezing. In section $4$ we discuss the evolution of
the Wigner function and  the purity for the single-mode case. In
section $5$ we give the main conclusions from the results.

\section{Model formalism and dynamical solution}

In this section we give the Hamiltonian for the system and derive
the solutions for its  equations of motion.

The Hamiltonian controlling the three-mode codirectional Kerr
nonlinear coupler can be represented as
\begin{eqnarray}
\begin{array}{lr}
\frac{\hat{H}}{\hbar }=\sum\limits_{j=1}^{3}(\omega _{j}\hat{a}^{\dagger }_j\hat{%
a}_j+\chi _{j}\hat{a}_{j}^{\dagger 2}\hat{a}_{j}^{2})+\overline{\chi }_1\hat{a}%
_{1}^{\dagger }\hat{a}_{1}\hat{a}_{2}^{\dagger }\hat{a}_{2}+\overline{\chi }_2\hat{a}%
_{1}^{\dagger }\hat{a}_{1}\hat{a}_{3}^{\dagger }\hat{a}_{3}+\overline{\chi }_3\hat{a}%
_{2}^{\dagger }\hat{a}_{2}\hat{a}_{3}^{\dagger }\hat{a}_{3}   \\
\\
+\lambda _{1}[\hat{a}_{1}\hat{a}_{2}^{\dagger }\exp (i\Delta_1 t)+
\hat{a}_{1}^{\dagger }\hat{a}_{2}\exp (-i\Delta_1 t)]
+\lambda _{2}[\hat{a}_{1}\hat{a}_{3}^{\dagger }\exp (i\Delta_2%
t)+\hat{a}_{1}^{\dagger }\hat{a}_{3}\exp (-i\Delta_2 t)],
\label{1}
\end{array}
\end{eqnarray}
%\end{document}
where $\Delta_1 =\omega _{1}-\omega _{2},$ and $\Delta_2=\omega
_{1}-\omega _{3}$. The waves are designated by $\hat{a}_1$
(fundamental and/or first), $\hat{a}_2$ (second) and $\hat{a}_3$ (third) modes
with frequencies $\omega_1, \omega_2$ and $\omega_3$,
respectively. The coupling constants $\chi_j$ and $\overline{\chi
}_j$ are proportional to the third-order susceptibility
$\chi^{(3)}$ and are responsible correspondingly for the
self-action and cross-action processes of the $j$th mode. The
linear coupling between the waveguides are represented by
$\lambda_1$ and $\lambda_2$, which we assume to be real.
Hamiltonian (\ref{1}) gives a generalization to several models discussed
in the literatures earlier by controlling the values of the interaction
parameters such as anhramonic oscillator \cite{anh}, up-conversion process
\cite{upc} and two-mode KNC \cite{faisal2}.
The
scheme describing Hamiltonian (\ref{1}) is shown in Fig. 1 and can
be explained as follows.  Two waveguides are operating by
Kerr-like nonlinear processes. In the first waveguide the
fundamental mode $\hat{a}_1$ propagates, while in the second
waveguide the second $\hat{a}_2$ and third $\hat{a}_3$ modes
propagate. The interaction between the fundamental mode and the
second-third modes occurs via the evanescent waves. We have
assumed that the all waves are propagating with the same velocity
$v$, hence the time $t$ and travelled distance $z$ are related by
$z=vt$. Outgoing fields can be detected as single or compound
modes by means of homodyne, photocounting or coincidence detection
in the standard way.
 Throughout the investigation of the system  we
do not consider  dissipation, which generally decreases the amount
of  nonclassical effects.

We have to comment that the linear coupling coefficient is from the wave equation equal
to $\lambda/v = \omega^2 \chi^{(2)}/2kc^2$, whereas the nonlinear
coupling coefficients are $\chi/v = \omega^2 \chi^{(3)}/2kc^2$
($v$ being the speed of light in the medium, $\omega$ being
frequency of the travelled wave and $k$ is its wave number).
Taking into account that values of quadratic susceptibilities lie
in an interval $10^{-14} - 10^{-10} m/V$ and values of cubic
susceptibilities in an interval $10^{-17} - 10^{-13} m^2/V^2$
for various nonlinear materials,  then in the optical region
the values of $\lambda/v$ lie from $10^4$ to $10^7 m^{-1}$  and values
of $\chi/v$ from $10$ to $10^5 m^{-1}$.  So the effects predicted should be
expected to be observable in samples of the length of centimeters
and in nanosecond scales.

Now introducing new operators  $\hat{A}_{j}= \hat{a}_{j}
\exp (i\omega _{j}t), j=1,2,3$ we can write the Heisenberg
equations for (\ref{1})  as
\begin{eqnarray}
\begin{array}{lr}
\frac{d\hat{A}_{1}}{dt} =-i(2\chi_1 \hat{A}_{1}^{\dagger
}\hat{A}_{1}+
\overline{\chi }_1\hat{A}_{2}^{\dagger }\hat{A}_{2}+\overline{\chi }_2\hat{A}%
_{3}^{\dagger }\hat{A}_{3})\hat{A}_{1}-i\lambda _{1}\hat{A}_{2}-i\lambda _{2}\hat{A}%
_{3}, \\
\\
\frac{d\hat{A}_{2}}{dt} =-i(2\chi_2 \hat{A}_{2}^{\dagger
}\hat{A}_{2}+
\overline{\chi }_1\hat{A}_{1}^{\dagger }\hat{A}_{1}+\overline{\chi }_3\hat{A}%
_{3}^{\dagger }\hat{A}_{3})\hat{A}_{2}-i\lambda _{1}\hat{A}_{1},\\
 \\
\frac{d\hat{A}_{3}}{dt}=-i(2\chi_3 \hat{A}_{3}^{\dagger
}\hat{A}_{3}+
\overline{\chi }_2\hat{A}_{1}^{\dagger }\hat{A}_{1}+\overline{\chi }_3\hat{A}%
_{2}^{\dagger }\hat{A}_{2})\hat{A}_{3}-i\lambda _{2}\hat{A}_{1}.
\label{5}
\end{array}
\end{eqnarray}
To solve (\ref{5})  exactly we assume
the processes of the self-action and cross-action compensate
each other in the evolution of the system.
This can be expressed as
 $\chi =\chi _{1}=\chi
_{2}=\chi _{3},\overline{\chi }=\overline{\chi }_1=\overline{\chi
}_2=\overline{\chi }_3 $ and $2\chi=\overline{\chi }$.
In fact, the cross-spectral coupling between various waveguides may be
comparable with the self-coupling provided that the surfaces of
the waveguides are of high quality (roughness is much less than
the wavelength) to avoid a substantional
reduction of amplitudes by evanescent damping. As follows from
the values for quadratic and cubic susceptibilites
both the linear and nonlinear couplings may be simultaneously
significant for sufficiently strong fields.
Under these conditions
one can easily prove that
%%%%%%%%%%%%%%%%%%%%%%%%%%%%%%%%%%%%%%%%%%%%%%%%%%%%%%%%%%%%%%%
\begin{figure}
\includegraphics[width=.80\linewidth]{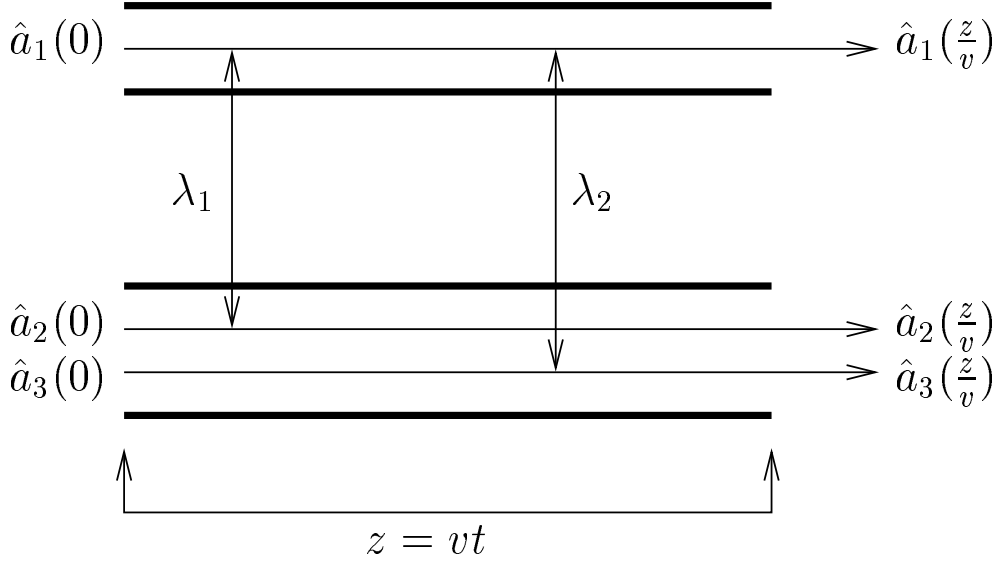}
\caption{ Scheme of realization of interaction (1).}
\end{figure}
%%%%%%%%%%%%%%%%%%%%%%%%%%%%%%%%%%%%%%%%%%%%%%%%%%%%%%%%%%%%

\begin{equation}
\hat{A}_{1}^{\dagger }\hat{A}_{1}+\hat{A}_{2}^{\dagger }\hat{A}_{2}+\hat{A}%
_{3}^{\dagger }\hat{A}_{3}=\hat{N}  \label{6}
\end{equation}
is a constant of motion. Based on this fact the system of
equations (\ref{5}) can be modified via  the substitution
$\hat{B}_j=\exp(2i\chi \hat{N}t)\hat{A}_j, j=1,2,3$ and takes the
form

\begin{eqnarray}
\begin{array}{lr}
\frac{d\hat{B}_{1}}{dt} = -i\lambda _{1}\hat{B}_{2}-i\lambda
_{2}\hat{B}
_{3}, \\
\\
\frac{d\hat{B}_{2}}{dt} = -i\lambda _{1}\hat{B}_{1}, \quad
\frac{d\hat{B}_{3}}{dt}= -i\lambda _{2}\hat{B}_{1}. \label{5a}
\end{array}
\end{eqnarray}
The system  (\ref{5a}) can be easily solved, e.g., using Laplace
transformation, and the general solution is
\begin{eqnarray}
\begin{array}{lr}
\hat{A}_{1}(t)=\exp (-2i\chi \hat{N}t)\Bigl\{ \hat{a}%
_{1}(0)\cos \mu t-i\frac{1}{\mu }(\lambda _{1}\hat{a}_{2}(0)+\lambda _{2}%
\hat{a}_{3}(0))\sin \mu t\Bigr\} , \\
\\
\hat{A}_{2}(t)=\exp (-2i\chi \hat{N}t)\Bigl\{ [1-2\frac{%
\lambda _{1}^{2}}{\mu ^{2}}\sin ^{2}(\frac{\mu
t}{2})]\hat{a}_{2}(0)-2\frac{\lambda _{1}\lambda _{2}
\hat{a}_{3}(0)}{\mu ^{2}}\sin ^{2}(\frac{\mu
t}{2})-i\frac{\hat{a}_{1}(0)\lambda
_{1}}{\mu }\sin( \mu t)\Bigr\},\\
 \\
\hat{A}_{3}(t)=\exp (-2i\chi \hat{N}t)\Bigl\{ [1-2\frac{%
\lambda _{2}^{2}}{\mu ^{2}}\sin ^{2}(\frac{\mu
t}{2})]\hat{a}_{3}(0)-2\frac{\lambda _{1}\lambda _{2}
\hat{a}_{2}(0)}{\mu ^{2}}\sin ^{2}(\frac{\mu t}{2})-i\frac{\lambda
_{2}\hat{a}_{1}(0) }{\mu }\sin (\mu t)\Bigr\},  \label{10}
\end{array}
\end{eqnarray}
where $\mu=\sqrt{\lambda_1^2+\lambda_2^2}$. There are several
facts which can be extracted from the structure of the Hamiltonian
(\ref{1}), the solution (\ref{10}) and Fig. 1. When
$\hat{a}_{3}(0)\leftrightarrow \hat{a}_{2}(0)$ and
$\lambda_1\leftrightarrow \lambda_{2}$ one obtains
$\hat{A}_{2}(t)\leftrightarrow \hat{A}_{3}(t)$. The nature of the
coupler, i.e. the switching of energy between waveguides,
manifests itself by periodic functions in (\ref{10}), whereas the
Kerr nonlinearities in the waveguides are described  by the
nonlinear (quadratic) phase, which plays an essential role in
generating the nonclassical effects. On the other hand,
 the fundamental mode $\hat{a}_1$ can  provide richer
nonclassical effects  than those produced by the second and third
modes as well as  the conventional coupler
\cite{faisal2,{faisal3}} (see Fig. 1). This is resulting from
 the coupler
mechanism, which switches  the energy
  in the second-third mode (second waveguide) jointly
to  the fundamental (first waveguide) and vice versa.

On the other hand, the nature of the field quantization
is evident where one  can easily prove from (\ref{10}) that
\begin{equation}
\left[\hat{A}_{j}(t),\hat{A}_{j'}^{\dagger }(t)\right] =\delta_{j,j'}, \label{hsh1}
\end{equation}
where $\delta_{j,j'}$ is the Kronecker delta and $j,j'=1,2,3$.  In the derivation
of (\ref{hsh1}) one has to use the identities

\begin{eqnarray}
\begin{array}{lr}
\exp (-2i\chi \hat{N}t)\hat{a}_{j}(0)\exp (2i\chi \hat{N}t) =
\hat{a}_j(0)\exp(2i\chi t), \\
\\
\exp (-2i\chi \hat{N}t)\hat{a}_{j}^{\dagger}(0)\exp (2i\chi \hat{N}t) =
\hat{a}^{\dagger}_j(0)\exp(-2i\chi t).
 \label{hsh2}
 \end{array}
\end{eqnarray}
Furthermore, the mean-photon number for the fundamental and second modes
are:
\begin{eqnarray}
\begin{array}{lr} \hat{A}_{1}^{\dagger }(t) \hat{A}_{1}(t)
= \hat{a}_{1}(0)\hat{a}^{\dagger}_{1}(0)\cos^{2} (\mu t)\\
\\
+\left[\lambda^{2} _{1}\hat{a}^{\dagger}_{2}(0)\hat{a}_{2}(0)
+\lambda^{2} _{2}
\hat{a}^{\dagger}_{3}(0)\hat{a}_{3}(0)+\lambda _{1}\lambda _{2}
\left(\hat{a}^{\dagger}_{2}(0)\hat{a}_{3}(0)
+\hat{a}_{2}(0)\hat{a}^{\dagger}_{3}(0)\right)\right]
\frac{\sin^{2} (\mu t)}{\mu^{2} }\\
\\
+i[
\lambda _{1}(\hat{a}_{1}(0) \hat{a}^{\dagger}_{2}(0)-
\hat{a}^{\dagger}_{1}(0) \hat{a}_{2}(0))
+\lambda _{2}(\hat{a}_{1}(0) \hat{a}^{\dagger}_{3}(0)-
\hat{a}^{\dagger}_{1}(0) \hat{a}_{3}(0))]
\frac{\sin (2\mu t)}{2\mu }\\
\\
\hat{A}_{2}^{\dagger }(t)
\hat{A}_{2}(t)
=
\frac{\lambda _{1}^{2}}{\mu ^{2}}\hat{a}^{\dagger}_{1}(0)\hat{a}_{1}(0)
\sin ^{2}(\mu t)
+ [1-2\frac{
\lambda _{1}^{2}}{\mu ^{2}}\hat{a}^{\dagger}_{2}(0)\hat{a}_{2}(0)
\sin ^{2}(\frac{\mu t}{2})]^{2}
+
4\frac{\lambda^{2} _{1}\lambda^{2} _{2}
}{\mu ^{4}}\hat{a}^{\dagger}_{3}(0)\hat{a}_{3}(0)
\sin ^{4}(\frac{\mu
t}{2})\\
\\
-
2\frac{\lambda _{1}\lambda _{2}
}{\mu ^{2}}
[1-2\frac{\lambda _{2}^{2}}{\mu ^{2}}\sin ^{2}(\frac{\mu
t}{2})]
\sin ^{2}(\frac{\mu t}{2})
[\hat{a}^{\dagger}_{2}(0)\hat{a}_{3}(0)
+\hat{a}_{2}(0)\hat{a}^{\dagger}_{3}(0)]
-
\frac{i\lambda _{1}}{\mu}
[1-2\frac{\lambda _{2}^{2}}{\mu ^{2}}\sin ^{2}(\frac{\mu
t}{2})]
\sin (\mu t)
\\
\\
\times [\hat{a}_{1}(0)\hat{a}^{\dagger}_{2}(0)
-\hat{a}^{\dagger}_{1}(0)\hat{a}_{2}(0)]
+\frac{2i\lambda^{2} _{1}\lambda _{2} }{\mu^{3}}
\sin ^{2}(\frac{\mu
t}{2})\sin (\mu t)
[\hat{a}_{1}(0)\hat{a}^{\dagger}_{3}(0)
-\hat{a}^{\dagger}_{1}(0)\hat{a}_{3}(0)].
  \label{hsh3}
\end{array}
\end{eqnarray}
From (\ref{hsh3}) the correlation between modes and switching mechanism,
 i.e. when one photon is created in
the one of the modes the other is annihilated in the other mode,
are quite obvious

We close this section by calculating the moments $ \langle
\prod\limits_{j=1}^{3}\hat{A}_{j}^{\dagger n_j}\hat{A}_{j}^{
m_j}\rangle$ when  the modes are initially prepared in the
coherent states $|\alpha_1,\alpha_2,\alpha_3\rangle$ and
$\alpha_j$ are real. From (\ref{10}) one can easily deduce that
\begin{equation}
 \langle \prod\limits_{j=1}^{3}\hat{A}_j^{\dagger n_j}\hat{A}_j^{
m_j}\rangle =\bar{\alpha}_1^{*n_1}(t)
\bar{\alpha}_1^{m_1}(t)\bar{\alpha}_2^{*n_2}(t)\bar{\alpha}_2^{m_2}(t)
\bar{\alpha}_3^{*n_3}(t)\bar{\alpha}_3^{m_3}(t)
z^{h_1}\exp[\epsilon(z^{h_2}-1)], \label{110}
\end{equation}
where $n_j, m_j$ are positive integers,
\begin{eqnarray}
\begin{array}{lr}
\epsilon=|\alpha_1|^2+|\alpha_2|^2+|\alpha_3|^2,\quad z=\exp(-2i\chi t),\\
\\
h_1=\frac{1}{2}\sum\limits_{j=1}^{3}[m_j(m_j-1)-n_j(n_j-1)]+
m_1(m_2+m_3)+m_2m_3-n_1(n_2+n_3)-n_2n_3,\\
\\
h_2=m_1+m_2+m_3-n_1-n_2-n_3 \label{2o}
\end{array}
\end{eqnarray}
and
\begin{eqnarray}
\begin{array}{lr} \bar{\alpha}_1(t)=\alpha_1\cos(\mu
t)-i[\lambda_1\alpha_2+\lambda_2\alpha_3]\frac{\sin(\mu t)}{\mu}
=\alpha_{1x}(t)+i\alpha_{1y}(t),\\
\\
\bar{\alpha}_2(t)=
\alpha_2[1-\frac{2\lambda_1^{2}}{\mu^2}\sin^{2}(\frac{1}{2}\mu t)]
-\frac{2\lambda_1\lambda_2\alpha_3}{\mu^2}\sin^{2}(\frac{\mu
t}{2}) -i\frac{\lambda_1\alpha_1}{\mu}\sin(\mu t)
=\alpha_{2x}(t)+i\alpha_{2y}(t),\\
\\
\bar{\alpha}_3(t)=
\alpha_3[1-\frac{2\lambda_2^{2}}{\mu^2}\sin^{2}(\frac{1}{2}\mu t)]
-\frac{2\lambda_1\lambda_2\alpha_2}{\mu^2}\sin^{2}(\frac{\mu
t}{2}) -i\frac{\lambda_2\alpha_1}{\mu}\sin(\mu t)
=\alpha_{3x}(t)+i\alpha_{3y}(t).  \label{3o}
\end{array}
\end{eqnarray}
Expression (\ref{110}) will be frequently used  in the paper. It
is obvious that when $\chi t=m\pi$ and $m$ is integer, i.e. $z=1$,
this expression  reduces to that for the coherent light with
amplitudes $\bar{\alpha}_j(t)$ indicating that the system produces
coherent light  periodically. Also from this expression one can
easily check that $\langle \hat{A}_j^{\dagger m_j}\hat{A}_j^{
m_j}\rangle = \langle \hat{A}_j^{\dagger }\hat{A}_j\rangle ^{
m_j}$. This means that the system is Poissonian, i.e. it cannot
exhibit sub-Poissonian statistics, and the mean-photon numbers
cannot provide RCP. Expression (\ref{110}) reduces to that of the
two-mode KNC \cite{faisal2} when $\lambda_2=0$ and $\alpha_3=0$.
Now we use the calculations given here  to investigate quadrature
squeezing, Wigner function and purity  in the following sections.
%%%%%%%%%%%%%%%%%%%%%%%%%%%%%%%%%%%%%%%%%
\section{Quadrature squeezing}
%%%%%%%%%%%%%%%%%%%%%%%%%%%%%%%%%%%%%%%%
Squeezing is a pure nonclassical phenomenon without classical
analog. Squeezed light has less noise in one of the field
quadrature than the vacuum level and an excess of noise in the
other  quadrature such that the uncertainty principle is
satisfied. Squeezed light can be measured in the homodyne
detection. This light has a lot of applications, e.g., in optical
communication networks \cite{yu}, quantum information \cite{inf1}
 and high precision measurements.
 In this section we investigate the single- and two-mode squeezing and shed the light
briefly on the evolution of the three-mode squeezing. In
doing so we define two quadratures $\hat{X}_n$ and $\hat{Y}_n$ as
\begin{equation}
 \hat{X}_n=\frac{1}{2}\sum\limits_{j=1}^{n} [\hat{A}_{j}(t)+
\hat{A}^{\dagger }_{j}(t)], \quad
 \hat{Y}_n=\frac{1}{2i}\sum\limits_{j=1}^{n}[\hat{A}_{j}(t)-
\hat{A}^{\dagger}_{j}(t)],  \label{ol5}
\end{equation}
where $n$ takes on values $1, 2$ and $3$ associated  with the
single-, two- and three-mode squeezing, respectively. Quadratures
(\ref{ol5}) satisfy the following commutation rule:
\begin{equation}
[\hat{X}_n,\hat{Y}_n]=\frac{iC_n}{2}, \label{ol1}
\end{equation}
where $C_n$ is a $c$-number and takes on the value  $C_1=1, C_2=2,
C_3=3$. The uncertainty relation associated with the commutation
rule (\ref{ol1}) is
\begin{equation}
\langle (\triangle \hat{X}_n)^{2}\rangle \langle (\triangle
\hat{Y}_n )^{2}\rangle \geq \frac{|C_n|^{2}}{16}, \label{ol2}
\end{equation}
where  $\langle (\triangle \hat{X}_n)^{2}\rangle=\langle
\hat{X}_n^{2}\rangle -\langle \hat{X}_n\rangle ^{2}$ and similar
form can be given for $\langle (\triangle \hat{Y}_n)^{2}\rangle$.
The system is said to be squeezed in the $X_n$-quadrature if
\begin{equation}
 S_n=4\langle (\triangle \hat{X}_n(t))^2\rangle -
|C_n|\leq 0. \label{ol3}
\end{equation}
The equality sign in (\ref{ol3}) holds  for minimum-uncertainty
states.
 Similar definition can be given for the $Y_n$-quadrature
(defining a $Q_n$-factor). Based on (\ref{110}) and (\ref{ol3}) we
investigate the single- and two-mode squeezing in details and comment on
the three-mode squeezing.
 As we deal with the Kerr media we can expect
squeezing in the guided modes in all these types.

 %%%%%%%%%%%%%%%%%%%%%%%%%%%%%%%%
\subsection{Single-mode squeezing}
%%%%%%%%%%%%%%%%%%%%%%%%%%%%%%%%%%
The squeezing factors for the $j$th mode can be straightforwardly
evaluated as
\begin{eqnarray}
\begin{array}{lr} S^{(j)}_1(t)= 2\alpha^{2}_{jx}(t)+
2\alpha^{2}_{jy}(t)+2\Bigl\{ [\alpha^{2}_{jx}(t)-
\alpha^{2}_{jy}(t)]\cos\Theta_1+2\alpha_{jx}(t)
\alpha_{jy}(t)\sin\Theta_1\Bigr\}\exp[-2\epsilon\sin^{2}(2\chi t)],\\
\\
-4\left[\alpha_{jx}(t)\cos\Theta_2 +
\alpha_{jy}(t)\sin\Theta_2\right]^{2}
\exp[-4\epsilon\sin^{2}(\chi t)],\\
\\
Q^{(j)}_1(t)= 2\alpha^{2}_{jx}(t)+ 2\alpha^{2}_{jy}(t)-2\Bigl\{
[\alpha^{2}_{jx}(t)-
\alpha^{2}_{jy}(t)]\cos\Theta_1+2\alpha_{jx}(t)
\alpha_{jy}(t)\sin\Theta_1\Bigr\}\exp[-2\epsilon\sin^{2}(2\chi t)]\\
\\
-4\left[\alpha_{jx}(t)\sin\Theta_2 -
\alpha_{jy}(t)\cos\Theta_2\right]^{2} \exp[-4\epsilon\sin^{2}(\chi
t)],  \label{fs1}
\end{array}
\end{eqnarray}
where
\begin{equation}
\Theta_1=2\chi t+\epsilon\sin(4\chi t),\qquad
\Theta_2=\epsilon\sin(2\chi t). \label{fs2}
\end{equation}
 The
subscript $1$ means  single-mode squeezing and the superscript $j$
denotes the mode under consideration. Now we give some analytical
results showing that the system can provide single-mode squeezing
depending  on the values of the interaction parameters. For
instance, when $(\alpha_1,\alpha_2,\alpha_3)=(\alpha,0,0)$ and
$\chi t=m\pi/2$, $m$ is odd integer, expressions (\ref{fs1}) give
for the fundamental and second modes the following:
\begin{eqnarray}
\begin{array}{lr} S^{(1)}_1(t)= -4\alpha^{2}\cos^{2}(\mu t)
\exp(-4\epsilon),\quad
Q^{(1)}_1(t)= 4\alpha^{2}\cos^{2}(\mu t),\\
\\
S^{(2)}_1(t)= 4
\frac{\lambda^{2}_1\alpha^{2}}{\mu^{2}}\sin^{2}(\mu t),\quad
Q^{(2)}_1(t)=
-4\frac{\lambda^{2}_1\alpha^{2}}{\mu^{2}}\sin^{2}(\mu
t)\exp(-4\epsilon). \label{fs3}
\end{array}
\end{eqnarray}
It is worth reminding that the expression associated with the
third mode can be obtained from that of the second mode via the
transformation $\lambda_1\leftrightarrow \lambda_2$ and
$\alpha_2\leftrightarrow \alpha_3$. From (\ref{fs3}) the
complementarity between the squeezing factors in the two
waveguides is obvious, i.e. maximum squeezing in the first
waveguide is accompanied by  minimum squeezing in the second
waveguide and vice versa. Maximum squeezing occurs in the
fundamental mode when $\mu t=m'\pi$ and $m'$ is integer. In this
case, $S^{(1)}_1(t)$ can be expressed as
\begin{equation}
S^{(1)}_1( t=m'\pi/\mu)=-\eta\exp(-\eta),\quad \eta=4\alpha^2.
\label{fsss1}
\end{equation}
By evaluating the extreme values for (\ref{fsss1})  one finds that
the maximum squeezing  in $S^{(1)}_1(t)$ is $37\%$ when
$\alpha=0.5$. Also  form (\ref{fs3}) it can be proved that the
maximum squeezing involved in $S^{(2)}_1(t)$ is the half of that
of $S^{(1)}_1(t)$, i.e. $18.5\%$. It is worth reminding that for the single-mode
case $C_1=1$, i.e. $S^{(j)}_1(t)$ is normalized.
Now we clarify the role of the
switching mechanism in the behaviour of the coupler since the
Kerr-like media can generate nonclassical effects regardless of
entanglement between modes. This can be understood by assuming
that $\lambda_1=\lambda_2=0$, i.e. the linear interaction between
waveguides is neglected, and $\alpha_1=0$. In this case the
fundamental mode cannot generate nonclassical effects, however, in
the framework of coupler  the fundamental mode  exhibits
nonclassical effects. In other words, entanglement between
different components of the system can generate as well as amplify
 the nonclassical effects (if they exist).
%%%%%%%%%%%%%%%%%%%%%%%%%%%%%%%%%%%%%%%%%%%%%%%%%%%%%%%%%%%%%%%
\begin{figure}
  \includegraphics[width=.80\linewidth]{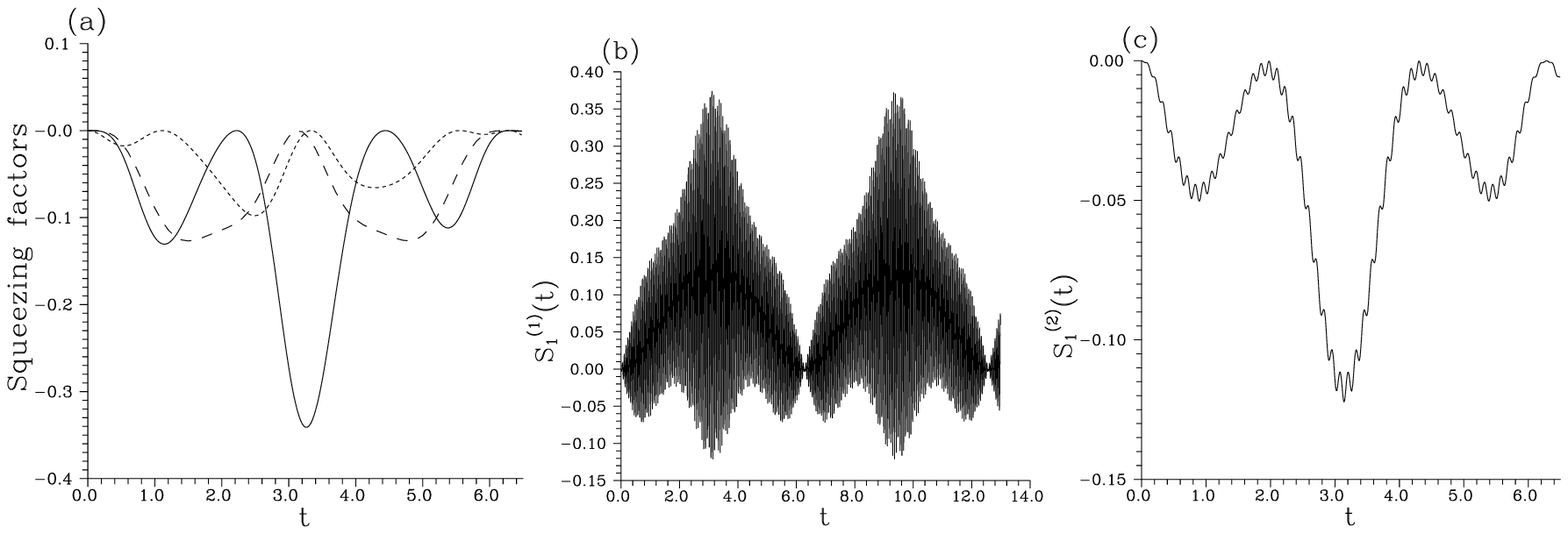}
   \caption{
Evolution of the single-mode squeezing factors for
 $\chi =0.5s^{-1}, (\alpha_2,\alpha_3)=(0.3,0.3)$ and for
(a) $(\alpha_1,\lambda_1,\lambda_2)=(0,1s^{-1},1s^{-1})$ (solid curve for
$Q_1^{(1)}(t)$, short-dashed curve for $S_1^{(2)}(t)$ and
long-dashed curve is the squeezing factor $Q_1^{(1)}(t)$ for the
two-mode KNC when $(\alpha_1,\alpha_2,\alpha_3,\lambda_1,\lambda_2)
=(0,0.3\sqrt{2}s^{-1}
,0,1s^{-1},0s^{-1})$), and
$(\alpha_1,\lambda_1,\lambda_2)=(0.3,1s^{-1},50s^{-1})$ for (b) $S_1^{(1)}(t)$
and (c) $S_1^{(2)}(t)$.}
\end{figure}
%%%%%%%%%%%%%%%%%%%%%%%%%%%%%%%%%%%%%%%%%%%%%%%%%%%%%%%%%%%%
Information about this situation is shown in  Fig. 2(a)
 for the given values of the
parameters. It is worth mentioning that
for $Q_1^{(1)}(t)$ of the two-mode KNC (as well as $S_2(t)$ in Fig. 3(a))
 the initial total photon number has been considered
 equal to that of the three-mode KNC for significant comparison.
 In this case we have noted that the squeezing occurs
only in the one of the quadratures, i.e. in $Y_n$
($X_n$)-quadrature for the fundamental (second) modes. Also
 squeezing generated
in the fundamental mode is much greater than that produced in the
second mode as well as in the single-mode of the two-mode KNC.
 Influence
of the values of the linear coupling constants $\lambda_j$, i.e.
the intensities of energy switching between the waveguides, on the
evolution of the quadrature squeezing is shown in Figs. 2(b) and
(c) for $S_1^{(1)}(t)$ and $S_1^{(2)}(t)$, respectively. It is
worth mentioning that $Q_1^{(1)}(t)$ provides a behaviour typical
to that shown in Fig. 2(b). From Fig. 2(b) one can see that
additionally to the occurrence of the nonclassical squeezing the
$S_1^{(1)}(t)$ exhibits leaf-revival-collapse phenomenon.
Actually, there are two requirements for occurring such
phenomenon, which are: (i) The  linear and nonlinear couplings
 in the Hamiltonian (\ref{1}) should contribute significantly to the dynamics of the system. (ii) The
energy exchange between the mode under consideration with at least
one of the modes in the  other waveguide is very strong. With this
in mind we can understand why $S_1^{(2)}(t)$ provides squeezing
rather than RCP. This can be realized easily  when
$\lambda_2>>\lambda_1$, i.e. $\lambda_1/\lambda_2\simeq 0$,
$\bar{\alpha}_2(t)=\alpha$ and hence the time evolution of
$S_1^{(2)}(t)$ basically depends on the contribution of the
Kerr-like medium, which alone cannot produce RCP. Generally, the
behaviour of the system for this case can be explained as follows.
As $\lambda_2>>\lambda_1$, i.e. the energy switching mainly occurs
between the fundamental and third modes, which behave as a
two-mode KNC and the second mode  practically does not affect the
behaviour.  Consequently,  the second mode behaves as an
independent mode evolving in a Kerr medium. In conclusion, when
$\lambda_1$ and $\lambda_2$ are large or $\lambda_1>>\lambda_2$
the squeezing factors for the fundamental and second modes provide
RCP. On the other hand, the occurrence of the RCP in the squeezing
factors of the fundamental mode can be explained in the following
sense \cite{faisal2}. The squeezing factors  (\ref{fs1}) include
two forms of periodic functions: one is coming from the
self-cross-nonlinear interaction part of (\ref{1}), in particular,
the exponential function whose period is $\pi/\chi$, and the other
is arising from the linear-interaction part whose period is
$\pi/\mu$. When the linear interaction between waveguides is very
strong--either between the fundamental-second or fundamental-third
or both--the period of the energy exchange between waveguides
decreases, i.e. many oscillations occur, till the interaction time
becomes $t=\pi/\chi$, at this moment the field is trapped
instantaneously by nonlinearity in the waveguides and the
squeezing factors show collapse.  The oscillations of the RCP are
caused by the linear coupling $\lambda_j$ and the envelope of the
revivals is caused by the nonlinearity in the system. As the
interaction proceeds the phenomenon is periodically repeated.

We have to stress that the evolution of the squeezing factors is
sensitive to the values of the intensities, however, in the
strong-intensity regime it can provide nonclassical squeezing
instantaneously greater than that present in Figs. 2, but the
behaviour becomes rather complicated. We draw the attention to
this point in the following section.

%%%%%%%%%%%%%%%%%%%%%%%%%%%%%%%%%%%%%%%%%%%%%%%%%%%%%%%%%%%%%%
\subsection{Two-mode squeezing}
%%%%%%%%%%%%%%%%%%%%%%%%%%%%%%%%%%%%%%%%%%%%%%%%%%%%%%%%%%%%%%

In this part we study the two-mode squeezing in which the
correlation between modes starts to play a role. Also we compare
the behaviour  here with the corresponding one for the two-mode
KNC. We restrict the analysis to the fundamental-second (between
waveguides) and second-third modes (in the same waveguide). The
fundamental-second mode squeezing factors can be straightforwardly
evaluated as
\begin{eqnarray}
\begin{array}{lr} S^{(1,2)}_2(t)= \frac{1}{2}[ S^{(1)}_1(t)+
S^{(2)}_1(t)]+ 2\Bigl\{ [\alpha_{1x}(t)\alpha_{2x}(t)-
\alpha_{1y}(t)\alpha_{2y}(t)]\cos\Theta_2
\\
\\
+ [\alpha_{2x}(t)\alpha_{1y}(t)+\alpha_{1x}(t)\alpha_{2y}(t)]
\sin\Theta_2\Bigr\}\exp[-2\epsilon\sin^{2}(2\chi t)] +
2[\alpha_{1x}(t)\alpha_{2x}(t)+\alpha_{1y}(t)\alpha_{2y}(t)]\\
\\
-4\left[\alpha_{1x}(t)\cos\Theta_1 +
\alpha_{1y}(t)\sin\Theta_1\right] \left[\alpha_{2x}(t)\cos\Theta_1
+ \alpha_{2y}(t)\sin\Theta_1\right] \exp[-4\epsilon\sin^{2}(\chi
t)],\\
\\
Q^{(1,2)}_2(t)= \frac{1}{2}[ Q^{(1)}_1(t)+ Q^{(2)}_1(t)]- 2\Bigl\{
[\alpha_{1x}(t)\alpha_{2x}(t)-
\alpha_{1y}(t)\alpha_{2y}(t)]\cos\Theta_2
\\
\\
+ [\alpha_{2x}(t)\alpha_{1y}(t)+\alpha_{1x}(t)\alpha_{2y}(t)]
\sin\Theta_2\Bigr\}\exp[-2\epsilon\sin^{2}(2\chi t)] +
2[\alpha_{1x}(t)\alpha_{2x}(t)+\alpha_{1y}(t)\alpha_{2y}(t)]\\
\\
-4\left[\alpha_{1y}(t)\cos\Theta_1 -
\alpha_{1x}(t)\sin\Theta_1\right] \left[\alpha_{2y}(t)\cos\Theta_1
- \alpha_{2x}(t)\sin\Theta_1\right] \exp[-4\epsilon\sin^{2}(\chi
t)],  \label{fs5}
\end{array}
\end{eqnarray}
%%%%%%%%%%%%%%%%%%%%%%%%%%%%%%%%%%%%%%%%%%%%%%%%%%%%%%%%%%%%%%%
\begin{figure}
  \includegraphics[width=.80\linewidth]{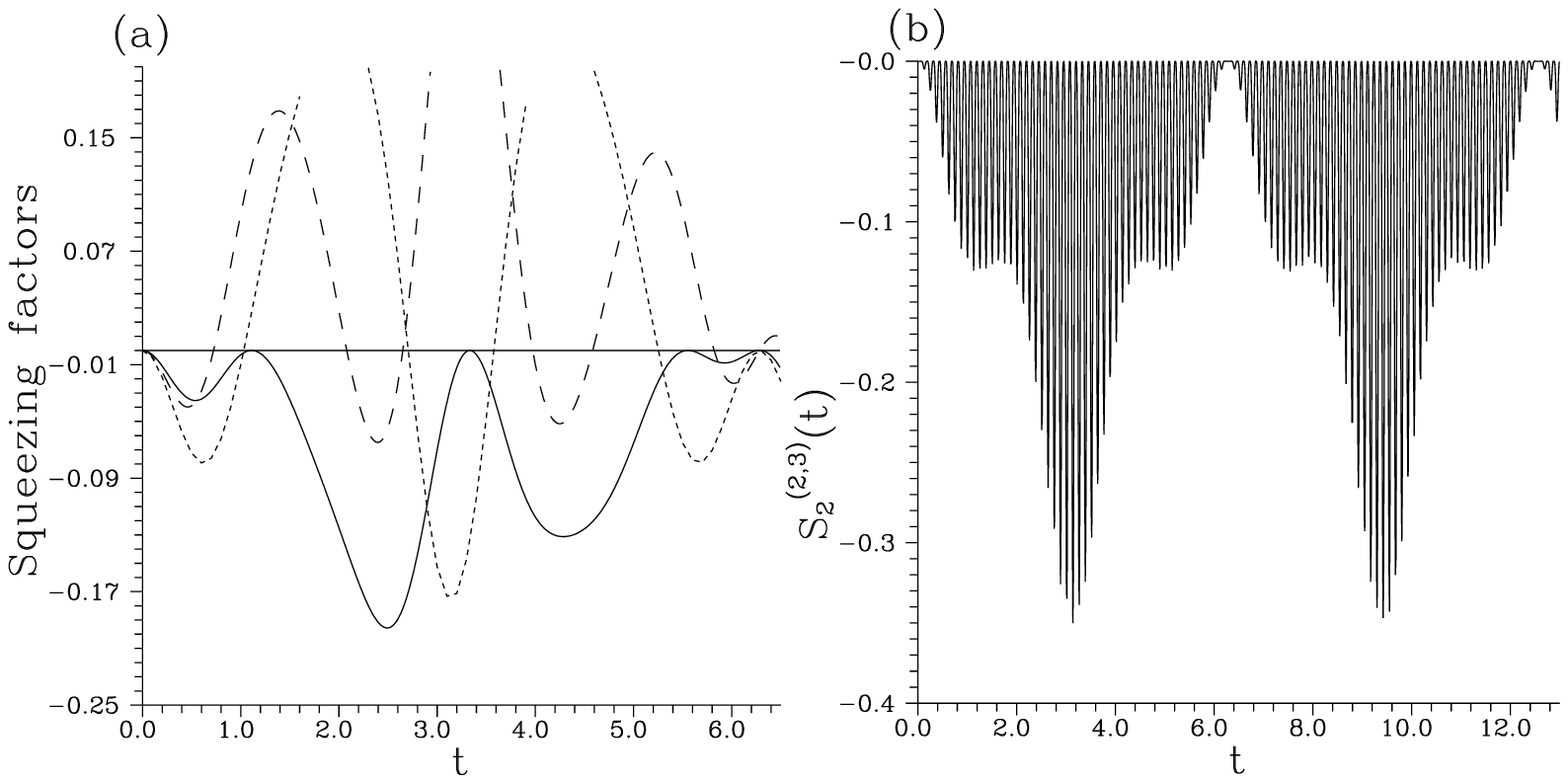} \caption{
Evolution of the two-mode squeezing factors for
 $ \chi =0.5s^{-1}, (\alpha_1,\alpha_2,\alpha_3)=(0,0.3,0.3)$ and for
(a) $\lambda_1=\lambda_1=1s^{-1}$ (long-dashed curve for
$S_1^{(1,2)}(t)$, solid curve for $S_2^{(2,3)}(t)$,  the
short-dashed curve is  given for $S_2(t)$  factor of the two-mode
KNC when
$(\alpha_1,\alpha_2,\alpha_3,\lambda_1,\lambda_2)=(0,0.3\sqrt{2},
0,1s^{-1},0s^{-1}))$ and the straight line is given for showing
the squeezing bound), (b) $S_2^{(2,3)}(t)$ for
$\lambda_1=50s^{-1},\lambda_2=1s^{-1}$.}
\end{figure}
%%%%%%%%%%%%%%%%%%%%%%%%%%%%%%%%%%%%%%%%%%%%%%%%%%%%%%%%%%%%
where $S_1^{(j)}$ and $Q_1^{(j)}$ are given by (\ref{fs1}) and
$\Theta_j$ have the same meaning as given above.
The subscript $2$ stands for two-mode squeezing and the
superscript $(1,2)$ denotes fundamental-second squeezing factor.
The second-third  squeezing factors can be easily obtain from
(\ref{fs5}) by simply replacing $1$ by $3$. As we did above we
analyse  the case $(\alpha_1,\alpha_2,\alpha_3)=(\alpha, 0,0)$ and
$\chi t=m\pi/2$, $m$ is odd integer, for which expressions
(\ref{fs5}) reduce to
\begin{eqnarray}
\begin{array}{lr} S^{(1,2)}_2(t)= 2\alpha^2[
\frac{\lambda^{2}_1}{\mu^2}\sin^2(\mu t)- \cos(\mu t)
\exp(-4\epsilon)],
\\
Q^{(1,2)}_2(t)= 2\alpha^2[\cos(\mu
t)-\frac{\lambda^{2}_1}{\mu^2}\sin^2(\mu t) \exp(-4\epsilon) ].
 \label{fs6}
\end{array}
\end{eqnarray}
Squeezing can be periodically realized in the two quadratures for
particular values of interaction time.  The two-mode squeezing
between the second and third modes, i.e. modes evolve in the same
waveguide, for this special case is

\begin{eqnarray}
\begin{array}{lr}
S^{(2,3)}_2(t)= 2\alpha^2\frac{(\lambda_1+\lambda_2)^2}{\mu^2}\sin^2(\mu t),\\
\\
 Q^{(2,3)}_2(t)=
-2\alpha^2\frac{(\lambda_1+\lambda_2)^2}{\mu^2}\sin^2(\mu t)
\exp(-4\epsilon). \label{fs7}
\end{array}
\end{eqnarray}
Squeezing can occur in the second quadrature only similar to the
single-mode case. Now we draw the attention to the behaviour of
the general forms (\ref{fs5}), which are presented in Figs. 3(a)
and (b) for given values of the interaction parameters.  From Fig.
3(a) one can observe that  $S^{(1,2)}_2(t)$  and $S_2(t)$ exhibit
nonclassical squeezing periodically.  Complementary behaviours
have been noticed for $Q^{(1,2)}_2(t)$ and $Q_2(t)$.
 Nevertheless,  for
modes evolving in the same waveguide where switching mechanism
does not exist directly, squeezing  occurs--for these values of
interaction parameters--only in one of the two quadratures, i.e.
$S^{(2,3)}_2(t)$, as indicated by the solid curve. We have to
stress that  the three-mode KNC provides amount of squeezing in
the framework of two-mode squeezing  greater than that of the
two-mode KNC (compare the solid curve with the short-dashed
curve). The action of the coupling constant is shown in Fig. 3(b)
for $S^{(2,3)}_2(t)$. From this figure it is obvious that
squeezing (or minimum-uncertainty light) always occurs and the
amount of squeezing is greater than that obtained for
weak-coupling case (compare with the solid curve in Fig.
3(a)). This means that the linear mechanism in the system can play
a significant
 role in amplifying the nonclassical effects. RCP is remarkable, which
always occurs provided that $\lambda_1$ or $\lambda_2$  or both
are large. Also for these cases we have noted that the two-mode
squeezing factors between the waveguides exhibit RCP somewhat
similar to that shown in Fig. 2(b).

Finally, for the three-mode squeezing factors we have noted that
 squeezing  periodically occurs in the two quadratures
 similar to that of the two-mode case, in particular,
the fundamental-second mode factors. Moreover, the amount of
squeezing achieved in the three-mode factors are greater
than those in the single-mode and two-mode
factors.
Also
 intense switching between the waveguides provide RCP in the evolution
 of the three-mode squeezing factors.

%%%%%%%%%%%%%%%%%%%%%%%%%%%%%%%%%%%%%%%%%%%%%%%%%%%%%
\section{Quasiprobability distribution function}
%%%%%%%%%%%%%%%%%%%%%%%%%%%%%%%%%%%%%%%%%%%%%%%%%%%%%
Quasiprobability distribution functions, namely, Wigner $W$-,
Husimi $Q$- and Glauber $P$-functions \cite{wign}, provide all
physical information about the quantum mechanical systems.
Investigation of these functions for the quantum mechanical
systems is one of the main topics in quantum optics. Actually,
among all quasiprobability functions $W$ function takes  a
considerable interest since it has been experimentally realized by
different techniques, e.g. in homodyne tomography \cite{tom},
photon counting experiment \cite{coun} and in  experiments with
trapped ions \cite{trap}, and it is sensitive to the interference
in phase space.
  In this section we investigate the single-mode
quasiprobability distribution functions, in particular, $W$
function for the system under consideration. As the derivations of
these functions are similar with those in \cite{faisal2}, we write
down only the explicit formulae for them. Just here we give only
outline of  the derivation. The solution (\ref{10}) includes
exponential quadratic operator phase, which causes difficulties in
 evaluating the quasiprobability functions in the standard way.
Thus one has to use the technique given in \cite{El-O1}, which
basically depends on the formula (\ref{110}), to achieve the goal.
It is worth reminding that the modes are initially prepared in the
coherent  state $|\alpha_1,\alpha_2,\alpha_3\rangle$. Furthermore,
we extend the investigation to include the purity for the
single-mode case, which can be evaluated by means of the symmetric
characteristic function. Now
  the $j$th mode
symmetric characteristic function is \cite{faisal2}

\begin{eqnarray}
\begin{array}{lr}
C_j(\zeta,t)=\exp(\frac{-1}{2}|\zeta|^{2})
\sum\limits_{n_{1},n_{2}=0}^{\infty}\frac{\zeta^{n_1}(-\zeta^{*})^{n_{2}}}
{n_1!n_2!}\bar{\alpha}^{n_2}_{j}(t) \bar{\alpha}^{*n_1}_{j}(t)\\
\\
\times z^{[\frac{n_2}{2}(n_2-1) -\frac{n_1}{2}(n_1-1)]}
\exp[\epsilon (z^{n_2-n_1}-1)]. \label{f6}
\end{array}
\end{eqnarray}

Also the $W$ function for the $j$th mode takes the form
\cite{faisal2}:

\begin{eqnarray}
\begin{array}{lr}
W_j(\beta,t)=\frac{2}{\pi }\exp(-2|\beta|^{2})
\sum\limits_{n_{1},n_{2}=0}^{\infty}\frac{(-1)^{n_1}\bar{\alpha}_{j}^{n_2}(t)
\bar{\alpha}_{j}^{* n_1}(t)}
{n_2!} 2^{n_2}\beta^{* n_2-n_1}\\
\\
\times z^{[\frac{n_2}{2}(n_2-1) -\frac{n_1}{2}(n_1-1)]}
\exp[\epsilon (z^{n_2-n_1}-1)] {\rm
L}_{n_1}^{n_2-n_1}(2|\beta|^{2}) , \label{f8}
\end{array}
\end{eqnarray}
where ${\rm L}^{m}_{n}(.)$ is the associated Laguerre polynomial
of order $n$ and $\beta=x+iy$. One can easily verify that when
$\chi t=m'\pi$ and $m'$ is integer, the
 system produces coherent light with amplitude
 $\bar{\alpha}_{j}(t)$ \cite{faisal2}.
%%%%%%%%%%%%%%%%%%%%%%%%%%%%%%%%%%%%%%%%%%%%%%%%%%%%%%%%%%%%%%%
\begin{figure}
 \includegraphics[width=.80\linewidth]{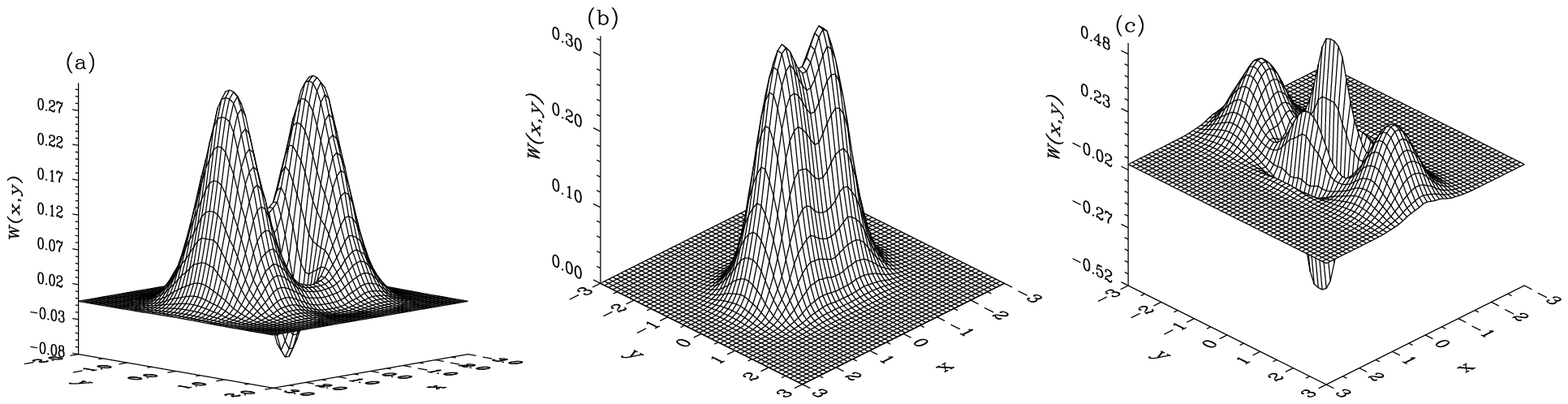}
\caption{The $W$ function for the single-mode case for $ \chi
=0.5s^{-1},t=\pi s$ and for  $
\alpha_j=0.9,\lambda_1=\lambda_2=1/(2\sqrt{2}s^{-1})$ (a)
(fundamental mode), (b) (second mode); and  $
\alpha_1=2,\alpha_2=\alpha_3=0$ and
$\lambda_1=\lambda_2=1/\sqrt{2}s^{-1}$ (c) (second mode).}
\end{figure}
%%%%%%%%%%%%%%%%%%%%%%%%%%%%%%%%%%%%%%%%%%%%%%%%%%%%%%%%%%%%
Furthermore, for $t\chi=(m'+1/2)\pi$ and $m'$ is integer,
expression (\ref{f8}) gives  that for the cat states (the
derivation is given in the appendix of \cite{faisal2})
\begin{eqnarray}
\begin{array}{lr} W_j(\beta,t)=\frac{1}{\pi} \Bigl\{
\exp\left(-2|\beta-i\bar{\alpha}_j(t)|^{2}\right) +
\exp\left(-2|\beta+i\bar{\alpha}_j(t)|^{2}\right)\\
\\
+2\exp\left[-2(|\beta|^{2}+D_j )\right]
\sin\left(2(\beta\bar{\alpha}_j^{*}(t)
+\beta^{*}\bar{\alpha}_j(t))\right)\Bigr\}, \label{im1}
\end{array}
\end{eqnarray}
where
\begin{equation}
D_j=\epsilon -|\bar{\alpha}_j(t)|^{2}. \label{im2}
\end{equation}
Roughly speaking, this form tends to  that of the YSCS \cite{yur}
in the $j$th mode when $D_j\simeq 0$ and $\bar{\alpha}_j(t)\neq
0$. In contrast to the two-mode KNC, the fundamental and second
(or third) modes of the three-mode KNC can provide different types
of cat states for certain choice of the interaction parameters.
Now we give some analytical facts related to $D_j$. There are
three different special cases, which can be discussed for
(\ref{im1}) based on the values of $\alpha_j$ where we assume
$\lambda_1=\lambda_2$.

 (i) When
$\alpha_j=\alpha, j=1,2,3$ for the fundamental and second modes,
we obtain
\begin{eqnarray}
\begin{array}{lr}
D_1=\alpha^2[1+\cos^2(\mu t)],\\
\\
D_2=\alpha^2[2+\frac{1}{2}\sin^2(\mu t)].
 \label{iim1}
\end{array}
\end{eqnarray}
Expressions (\ref{iim1}) show that $D_2\geq D_1$ and $D_j\simeq 0$
only when $\alpha <1$. This means that the cat states can be
generated in the first waveguide faster than that in the second
waveguide. As  $\alpha$ is relatively small  the microscopic cat
states \cite{El-O1} can be generated in the two waveguides
simultaneously. Trivially, statistical-mixture coherent states can
be generated for this case in dependence on the values of the
interaction parameters.

(ii) For $(\alpha_1,\alpha_2,\alpha_3)=(\alpha,0,0)$ we obtain

\begin{eqnarray}
\begin{array}{lr}
D_1=\alpha^2\sin^{2}(\mu t),\\
\\
D_2=\alpha^2[1-\frac{1}{2}\sin^2(\mu t)].
 \label{iim2}
\end{array}
\end{eqnarray}
 Expressions (\ref{iim2}) indicate that
when the fundamental mode provides YSCS  in a macroscopical form
$\alpha>1$, say, when $\mu t=\pi$, the second (or the third)
produces statistical-mixture coherent states, however, when
$\alpha<1$, YSCS in the microscopic forms are simultaneously
generated in the two waveguides. This shows  the crucial role for
the intensities of the initial light.

(iii) For $(\alpha_1,\alpha_2,\alpha_3)=(0,\alpha,0)$ we obtain

\begin{eqnarray}
\begin{array}{lr}
D_1=\alpha^2[1-\frac{1}{2}\sin^2(\mu t)],\\
\\
D_2=2\alpha^2[1-\frac{1}{2}\sin^2(\frac{\mu}{2}
t)]\sin^2(\frac{\mu}{2} t).
 \label{iim3}
\end{array}
\end{eqnarray}
This case represents  the inverse of the case (ii). The above
cases lead to an important fact: In order to obtain
macroscopically distinguishable YSCS from a particular mode it
should be initially prepared in coherent state with large
intensity. Information on the above cases is shown in Figs. 4 for
the fundamental and second modes for given values of interaction
parameters. The values of the linear constants $\lambda_j$ have
been chosen to minimize $D_1$. In Fig. 4(a) for the fundamental
mode one can observe the two Gaussian peaks and inverted negative
peak inbetween indicating the occurrence of the interference in
phase space. Fig. 4(b) (for the second mode) provides information
on the statistical-mixture coherent states, i.e. it shows two-peak
forms without interference in phase space. We have to stress that
in Figs. 4(a) and (b) the states generated are close to those of
the microscopic regime. Fig. 4(c) gives the well-known shape for
the $W$ function of the macroscopically distinguishable YSCS. In
this case the fundamental mode collapses to the state:
\begin{equation}
|\psi_1(t)\rangle=\frac{1}{\sqrt{2}}[|i\bar{\alpha}_{1}(t)\rangle+\exp(i\frac{\pi}{2})
|-i\bar{\alpha}_{1}(t)\rangle]. \label{imm2}
\end{equation}
The $W$ function for the second mode corresponding to Fig. 4(c) is
that of the vacuum state, where $\bar{\alpha}_2(t)\simeq 0$.
Actually, we have noted that such behaviours of the fundamental
and second modes--related to Fig. 4(c) and $W$ function of vacuum
light--can be interchanged  when one uses $t=\pi s,
\alpha_1=\alpha_3=0, \alpha_2=2, \lambda_1=\lambda_2=\sqrt{2}s^{-1}$.

In Figs. 5 we illustrate the general case, when the above special
cases are no longer applicable. In these figures we consider
$t=12.38s$, where   squeezing is well pronounced in the behaviours
of the single-mode case for the fundamental and second modes. In
these figures the nonclassical effects are presented as asymmetric
multi-peak structure indicating generation of the multi-component
cat states. Also Fig. 5(a) includes additionally negative values
showing that the fundamental mode gives amount of the nonclassical
behaviour greater than that for the second (or third) mode.
 This agrees with the explanation
given in section 2. On the other hand, when $\lambda_1>>\lambda_2$
these negative values  disappear and the $W$ functions of the
fundamental and second modes are almost the same (see Fig. 5(c)).
It is worthwhile mentioning that the corresponding $W$ function
for the two-mode KNC gives two symmetric peaks without involving
any negative values, which is insensitive to  the values of
$\lambda_1 (\geq 1)$. In conclusion, the system can generate
different types of YSCS depending  on the values of the
interaction parameters.

%%%%%%%%%%%%%%%%%%%%%%%%%%%%%%%%%%%%%%%%%%%%%%%%%%%%%%%%%%%%%%%
\begin{figure}
 \includegraphics[width=.80\linewidth]{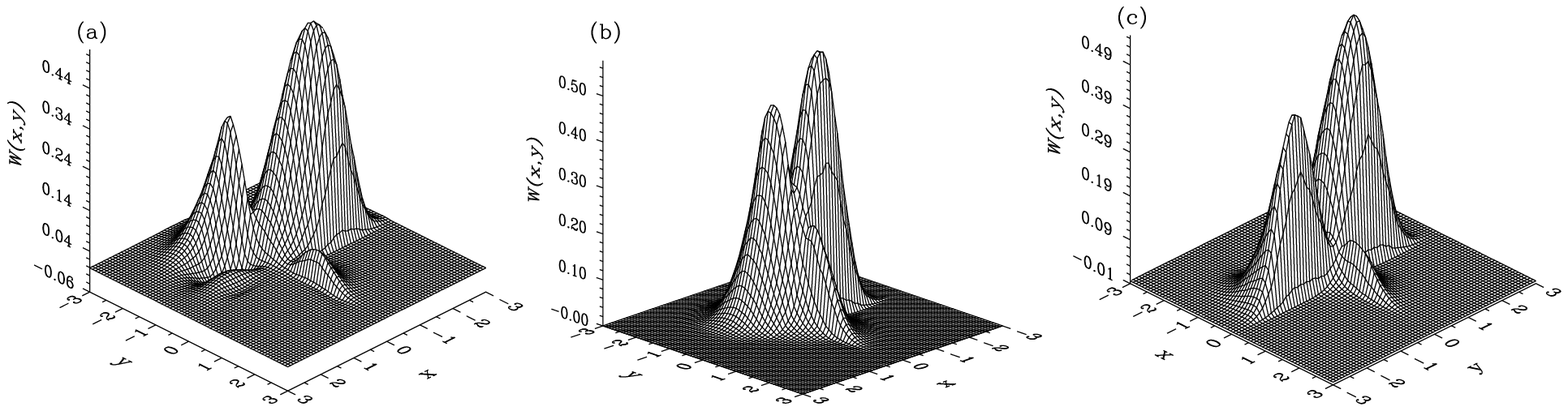} \caption{The $W$ function
for $ \chi =0.5s^{-1},t=12.38s,
\alpha_j=1,\lambda_1=\lambda_2=1s^{-1}$  and for (a) the
fundamental  mode, (b) second mode. Figure (c) is given for the
fundamental mode for the case $\lambda_1=50s^{-1},
\lambda_2=1s^{-1}$, the values of the others parameters are the
same as for (a).}
\end{figure}
%%%%%%%%%%%%%%%%%%%%%%%%%%%%%%%%%%%%%%%%%%%%%%%%%%%%%%%%%%%%
Entanglement characterizes intrinsically quantum-mechanical
correlations between quantum systems. In this regard quantum
entanglement is the basic resource required to implement several
of the most important processes studied by quantum information
theory. Thus, we discuss the single-mode degree of purity ${\rm
Tr}\hat{\rho}^{2}_j(t)$ for the system under consideration, which
gives globally information about the entanglement. Purity can be
determined by the knowledge of quantum state of the system, which
in turn can be obtained by quantum tomography \cite{tom}. Also it
is worth mentioning that in \cite{adam1}  entanglement has been
studied for the two-mode KNC driven by external classical field
showing that the system is able to generate Bell-like states. Also
in \cite{adam2} the purity and the relative entropy are
investigated for the pumped dissipative nonlinear oscillator
including Kerr nonlinearity. The purity  can be evaluated by means
of the symmetric characteristic function through  the relation

\begin{equation}
{\rm Tr}\hat{\rho}^{2}_j(t)=\frac{1}{\pi} \int |C_j(\zeta,t)|^{2}
d^{2}\zeta. \label{fff7}
\end{equation}
When ${\rm Tr}\hat{\rho}^{2}_j(t)=1$ the mode under consideration
 is in a  pure state, i.e. the mode under consideration is disentangled from the rest
of the system, while when ${\rm Tr}\hat{\rho}^{2}_j(t)<1$  the
mode is in a mixed state and consequently it is entangled with the
rest of the system. The  subsystems are most entangled when their
reduced density matrices are maximally mixed.
 Substituting
(\ref{f6}) into (\ref{fff7}) and after lengthy but straightforward
calculation we arrive at

\begin{eqnarray}
\begin{array}{lr} {\rm Tr}\hat{\rho}^{2}_j(t)=
\sum\limits_{n_{1},m_{1}=0}^{\infty}\frac{ (n_1+m_1)!
 (-1)^{m_1+n_1}|\bar{\alpha}_j(t)|^{2(n_1+m_1)} }{[n_1!m_1!]^2}
\\
\\
+ 2 \sum\limits_{m_{1}>m_{2}}^{\infty}
\sum\limits_{n_{1}=0}^{\infty}\frac{(n_1+m_1)!(-1)^{n_1+m_2}
|\bar{\alpha}_j(t)|^{2(n_1+m_1)}
}{n_1!m_1!m_2!(n_1+m_1-m_2)!}\exp[-
4\epsilon\sin^{2}\left((m_2-m_1)\chi
t\right)]\cos\psi(n_1,m_1,m_2), \label{f6f}
\end{array}
\end{eqnarray}
where
\begin{equation}
 \psi(n_1,m_1,m_2)= 2\chi t (m_2-m_1)
  (m_2-n_1). \label{f7f}
\end{equation}
It is evident that (\ref{f6f}) is dependent on the mean-photon
number for the mode under consideration. Thus for fixed value of
$\chi$ and $\chi t\neq m\pi, \quad m $ is integer,
 the evolution of the purity reflects well the switching
mechanism of the coupler provided that at least one of the initial
intensities is large.
%%%%%%%%%%%%%%%%%%%%%%%%%%%%%%%%%%%%%%%%%%%%%%%%%%%%%%%%%%%%%%%
\begin{figure}
\includegraphics[width=.80\linewidth]{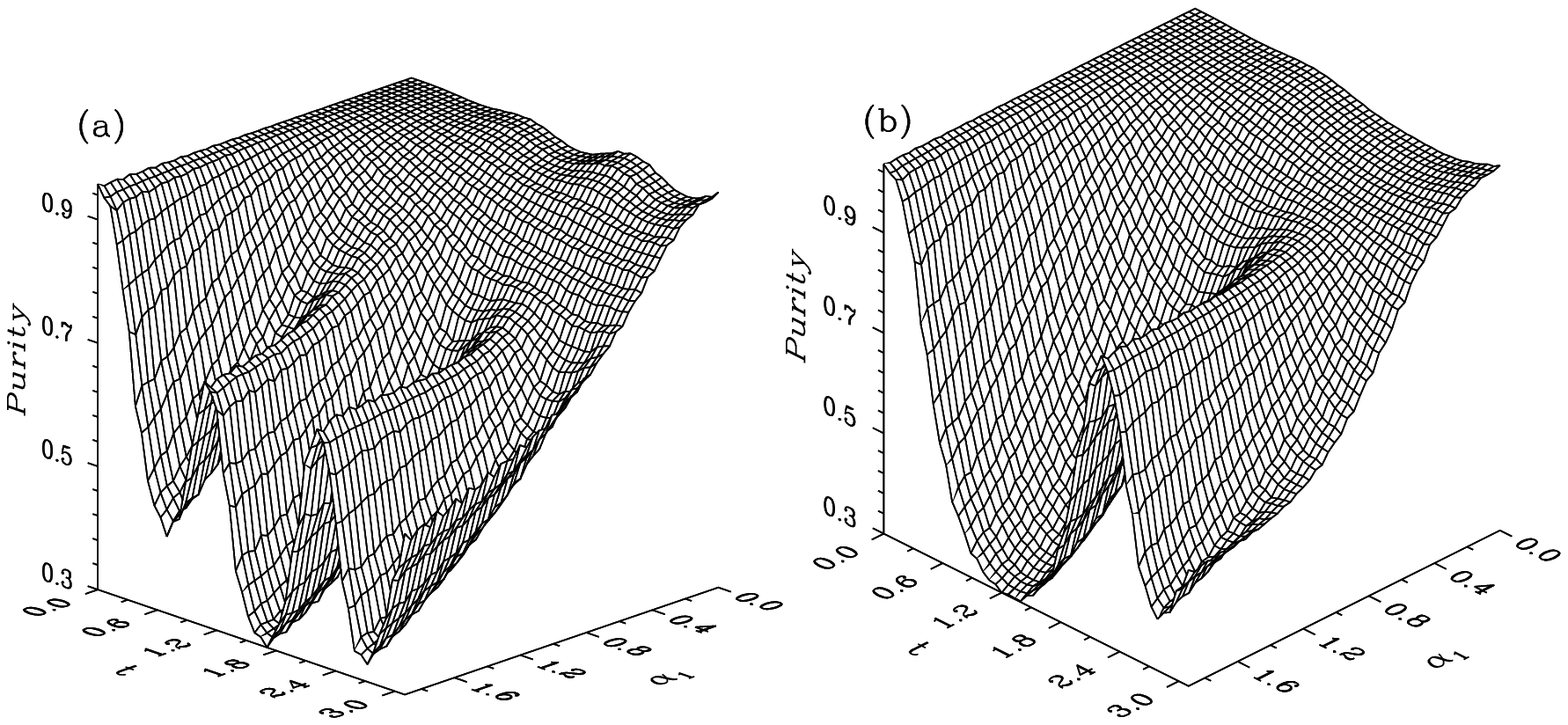}
   \caption{The evolution of the purity ${\rm Tr}\hat{\rho}^{2}_j(t)$
for $\chi =0.5s^{-1}, \alpha_1=\alpha_2=0.3$ and $\lambda_1=\lambda_2=1s^{-1}$  for the
fundamental mode (a) and the second mode (b).}
\end{figure}
%%%%%%%%%%%%%%%%%%%%%%%%%%%%%%%%%%%%%%%%%%%%%%%%%%%%%%%%%%%%
In relation to Figs. 6, i.e. $\lambda_1=\lambda_2,
\alpha_2=\alpha_3=\alpha$, the mean-photon numbers for the
fundamental and second modes take the forms
\begin{equation}
|\bar{\alpha}_1(t)|^{2}=\alpha_1^2\cos^2(\mu
t)+2\alpha^2\sin^{2}(\mu t),\quad
|\bar{\alpha}_2(t)|^{2}=\alpha^2\cos^2(\mu
t)+\frac{\alpha_1^2}{2}\sin^{2}(\mu t). \label{mod1}
\end{equation}
 When
$\alpha_j$ are small, the mean-photon numbers of the fundamental
and second modes would be relatively small and hence the evolution
of the purity will be close to that of the initial case (see Figs.
6(a) and (b) for given values of the parameters). Also from Figs.
6 we  note that the fundamental and second modes tend to partial
pure states, i.e. close to pure states, when their mean-photon
numbers provide their initial values. In this case the components
of the system are approximately disentangled. Additionally, the
fundamental mode  only   tends to almost pure state when the
energies in the two waveguides are completely interchanged. More
illustratively, from these figures one can observe that when $t=0s$
the two modes are in pure states. When $t$ increases and
$\alpha_1$ is large, entanglement between modes through the
interaction (\ref{1}) is achieved and both the fundamental and
second modes become mixed. The fundamental mode provides first
maximum mixedness at $t \simeq 0.69s$ (see Fig. 6(a)). Actually, at
this value of the interaction time, i.e. $t \simeq 0.69s$, the
mean-photon numbers for the three modes are equal and consequently
 their purities are equal, i.e. ${\rm Tr}\hat{\rho}^{2}_j(t)\simeq 0.37$
  (we have checked these facts).  When the interaction proceeds, switching energy
between waveguides continues until $t \mu =\pi/2$ then the
energies in the waveguides are completely interchanged (cf.
(\ref{mod1})). At this moment the fundamental (second) mode is  in
partial pure (maximum mixed) state since  its mean-photon number
is  small (large). This behaviour is repeated until $\mu t=\pi$
where the fundamental and second modes provide their initial
mean-photon numbers.

Now we draw the attention to find  analytically the values of the
interaction time for
 which the modes of the system are completely disentangled,
  i.e. ${\rm Tr}\hat{\rho}^{2}_j(t)=1$. This can be realized  from
 (\ref{110}) since all calculations in the paper depend on this
 expression. Therefore the condition for disentanglement is that
\begin{equation}
 \langle \prod\limits_{j=1}^{3}\hat{A}_j^{\dagger
n_j}\hat{A}_j^{ m_j}\rangle =
  \prod\limits_{j=1}^{3}\langle\hat{A}_j^{\dagger
n_j}\hat{A}_j^{ m_j}\rangle. \label{101}
\end{equation}
From the analysis given in the paper  this  occurs when $\chi
t=m\pi, \quad m $ is integer. In this case the modes are
completely disentangled and each of which has a coherent light
with time-dependent amplitude. On the other hand, the values of
interaction times for which the system provides its initial light
can be obtained by solving simultaneously the following two
equations
\begin{equation}
\mu t=2m'\pi,\quad \chi t=m\pi, \label{103}
\end{equation}
where $ m'$ and $ m$ are integers.
 These equations are connected with the requirements that
$\bar{\alpha}_j(t)=\alpha_j$. Equations (\ref{103}) lead to the
fact that  the system tends to the initial stage  when $\mu=\chi$
and  $\mu t=2m'\pi$. In this case the linearity of the system
compensates the nonlinearity.

%%%%%%%%%%%%%%%%%%%%%%%%%%%%%%%%%%%%%%%%%%%%%%%%%%%%
\section{Conclusion}
%%%%%%%%%%%%%%%%%%%%%%%%%%%%%%%%%%%%%%%%%%%%%%%%%%%%
In this paper we have discussed the  quantum properties for the
three-mode codirectional nonlinear Kerr coupler, when the modes
are initially prepared in coherent light. After obtaining the
exact solution of equations of motion we have investigated
single- and two-mode quadrature squeezing, $W$ function
and purity. Various interesting effects have been obtained.
Different forms of squeezing  have been achieved by varying the
parameters $\chi, \lambda_j$ and $\alpha_j$. Squeezing can be
equally shared between the guided modes and the interacting modes
can behave as two separated dynamical systems by controlling the
intensity of switching between waveguides. Quadrature squeezing
can exhibit leaf-revival-collapse phenomenon based on the
competition between the linearity and nonlinearity in the system.
Moreover, different forms of cat states have been generated and
confirmed in the evolution of $W$ function. We have analytically
proved that two different types of cat states can be
simultaneously realized in the evolution of the fundamental and
second (third) modes. Also we have discussed the mixedness for the
single-mode case and deduced
 the values of the interaction time as well as the conditions
required for complete disentanglement between the components of
the  system. The nature of the coupler  has manifested itself in
all studied quantities as complementary between the behaviours of
the fundamental and second-third modes, i.e. as switching energy
between waveguides. Moreover, we have showed that the nonclassical
effects  provided by the  fundamental mode are richer than those
for both the second and  third modes as well as for the two-mode
KNC. Finally, the system discussed in this paper is more effective
than the conventional Kerr coupler and can be used to amplify the
nonclassical effects.

\section*{Acknowledgement}
J.P. and F.A.A.E. thank the partial support from the grant
LN00A015 of the Czech Ministry of Education and from the EU
Project COST OCP 11.003.

%%%%%%%%%%%%%%%%%%%%%%%%%%%%%%%%%%%%%%%%%%%%%%%%%%%%%%%%%%%%%%%%%%%%%%%%%%%%%%%%
%%%%%%%%%%%%%%%%%%%%%%%%%%%%%%%%%%%%%%%%%%%%%%%%%%%%%%%%%%%%%%%%%%%%%%%%%%%%%%%%
\section*{References}
%%%%%%%%%%%%%%%%%%%%%%%%%%%%%%%%%%%%%%%%%%%%%%%%%%%%%%%%%%%%%%%%%%%%%%%%%%%%%%%%
%%%%%%%%%%%%%%%%%%%%%%%%%%%%%%%%%%%%%%%%%%%%%%%%%%%%%%%%%%%%%%%%%%%%%%%%%%%%%%%%

\end{document}